\input harvmac
\def\Title#1#2#3{#3\hfill\break \vskip -0.35in
\rightline{#1}\ifx\answ\bigans\nopagenumbers\pageno0\vskip.2in
\else\pageno1\vskip.2in\fi \centerline{\titlefont #2}\vskip .1in}


\def\bra#1{\langle #1 |}
\def\ket#1{| #1\rangle}

\def\S{{\cal{S}}}

\def\R{\hbox{\rm I \kern-5pt R}}

\font\ticp=cmcsc10
\def\ajou#1&#2(#3){\ \sl#1\bf#2\rm(19#3)}
%
%

\lref\ll{L.D.~Landau and E.M.~Lifschitz, {\sl ``Quantum Mechanics (
Non-relativistic Theory)''}, Third Edition (1977) Pergamon Press.}

\lref\diosi{L. Di\'osi "On maximum number of decoherent histories,"
gr-qc/9409028.}

\lref\griff{R.B.~Griffiths, \ajou J. Stat. Phys. &36 (84) 219.}

\lref\grifflogic{R.B.~Griffiths, \ajou Found. Phys. &23 (93) 1601.}

\lref\griffincon{Section 6.2 of \griff.}

\lref\grifftwo{Section 4.3 of \griff.}

\lref\omnes{R.~Omn\`es, \ajou J. Stat. Phys. &53 (88) 893;
\ajou ibid &53 (88) 933; \ajou ibid &53 (88) 957; \ajou
ibid &57 (89) 357.}

\lref\omnesreview{R.~Omn\`es, \ajou Rev.~Mod.~Phys. &64 (92) 339.}

\lref\omnestrue{Page 366 of \omnesreview.}

\lref\omnesmeasure{Page 362 of \omnesreview.}

\lref\omnesclass{Sections D-F of \omnesreview.}

\lref\omnesnew{R.~Omn\`es, \ajou Phys. Lett. A &187 (94) 26.}

\lref\gmhsantafe{M.~Gell-Mann and J.B.~Hartle in {\it Complexity, Entropy,
and the Physics of Information, SFI Studies in the Sciences of
Complexity}, Vol.
VIII, ed. by W.~Zurek,  Addison Wesley, Reading (1990).}

\lref\dowkerkent{F.~Dowker and A.~Kent, {\it ``On the Consistent
Histories Approach to Quantum Mechanics''}, J.~Stat.~Phys, to appear.} 

\lref\halliwell{J.~Halliwell, Imperial College preprint IC/93-94/52,
to appear in {\it Fundamental Problems in Quantum Theory}, ed. by
D. Greenberger}

\lref\gmhthree{M.~Gell-Mann and J.B.~Hartle in {\it Proceedings of the
NATO Workshop on the Physical Origins of Time Asymmetry, Mazag\'on, Spain,
September 30-October 4, 1991} ed. by J.~Halliwell, J.~P\'erez-Mercader, and
W.~Zurek, Cambridge University Press, Cambridge (1994), gr-qc/9304023.}

\lref\gmhprd{M.~Gell-Mann and J.B.~Hartle, \ajou Phys. Rev. D
&47 (93) 3345.}

\lref\unruhzurek{W.~Unruh and W.~Zurek, \ajou Phys. Rev. D &40 (89)
1071.} 

\lref\gmhonmeasure{Pages 448-449 of \gmhsantafe.}

\lref\gmhonbranches{Pages 450-451 of \gmhsantafe; page 3352 of \gmhprd.}

\lref\gmhpathint{Pages 432-434 of \gmhsantafe.}

\lref\gmhquasi{Page 445 of \gmhsantafe.}

\lref\gmhigus{Page 454 of \gmhsantafe.}

\lref\gmhaltern{Page 455 of \gmhsantafe.}

\lref\gmhtriv{Page 441 of \gmhsantafe.}

\lref\gmhcomm{M.~Gell-Mann and J.B.~Hartle, {\it Equivalent Sets of
Histories and Multiple Quasiclassical Domains},  Preprint UCSBTH-94-09,
gr-qc/9404013.}

\lref\gmhoverlap{Page 16 of \gmhcomm.}

\lref\hartleone{J.B.~Hartle, in {\it Quantum Cosmology and Baby
Universes}, Proceedings of the 1989 Jerusalem Winter School on
Theoretical Physics,
ed.~by S.~Coleman, J.~Hartle, T.~Piran, and S.~Weinberg,
World Scientific, Singapore (1991).}

\lref\hartletwo{J.B.~Hartle, \ajou Phys. Rev. &D44 (91) 3173.}

\lref\hartlethree{J.B.~Hartle, {\it Spacetime Quantum Mechanics and the
Quantum Mechanics of Spacetime} in {\it Gravitation and Quantizations:
Proceedings of the 1992 Les Houches
Summer School}, ed. by B.~Julia and J.~Zinn-Justin, North Holland
Publishing Co, Amsterdam, (1994), gr-qc/9304006.}

\lref\hartlepast{Page 77 of \hartleone.}

\lref\hartleequiv{Pages 25-26 of \hartleone.}

\lref\hartlemarvellous{Section V.1.2 of \hartleone.}

\lref\hartlepriv{J.B.~Hartle, private communication.}

\lref\omnespriv{R.~Omn\`es, private communication.}

\lref\griffithspriv{R.~Griffiths, private communication.}

\lref\isham{C.J.~Isham,
Lectures on Time and Quantum Gravity, St. Andrews (1993);
{\it Quantum Logic and the Histories Approach to Quantum Theory}
Imperial College preprint TP/92-93/39, gr-qc/9308006;
C.J.~Isham and N.~Linden,
 {\it Quantum temporal logic
and decoherence functionals in the histories approach to
generalised quantum theory} Imperial College
preprint TP/93-94/35, gr-qc/9405029.}

\lref\penrose{R.~Penrose, {\it ``The Nature of Space and Time''}, Isaac
Newton Institute debate (with S.W.~Hawking), May 1994.}

\lref\penrosebook{R.~Penrose, {\it ``Shadows of the Mind''}, Oxford
University Press, in press.}

\lref\mwi{ B.~DeWitt and R.N.~Graham,  eds. {\it The Many Worlds
Interpretation
of Quantum Mechanics}, Princeton University Press, Princeton (1973).}

\lref\everett{H.~Everett, \ajou Rev. Mod. Phys. &29 (57) 454.}

\lref\bell{J.S.~Bell, {\it ``Quantum Mechanics for Cosmologists''}, in
{\it ``Quantum Gravity 2''}, eds. C.~Isham, R.~Penrose and D.~Sciama,
Clarendon Press, Oxford (1981) 611-637.}

\lref\page{D.~Page, \ajou Phys. Rev. Lett. &70 (93) 4034.}

\lref\pagegoldstein{S.~Goldstein and D.~Page, {\it Linearly Positive
Histories},
 Alberta preprint Thy-43-93, gr-qc/9403055.}

\lref\jooszeh{E.~Joos  and H.D.~Zeh, \ajou Zeit. Phys. &B59 (85) 223.}

\lref\zeh{ H.D.~Zeh, \ajou Found. Phys. &1 (71) 69.}

\lref\zurek{W.~Zurek, \ajou Phys. Rev. &D24 (81) 1516;
\ajou Phys. Rev. &D26 (82) 1862.}

\lref\caldlegg{A.~Caldeira  and A.~Leggett, \ajou Physica &121A (83) 587.}

\lref\bellmwicrit{J.S.~Bell, {\it The Measurement Theory of Everett and
De Broglie's Pilot Wave}
in {\it Quantum Mechanics, Determinism, Causality and Particles}, M.~Flato
et al. (eds.), D.~Reidel, Dordrecht, 1976; reprinted in
J.S.~Bell, {\it Speakable and Unspeakable
in Quantum Mechanics}, Cambridge University Press, Cambridge (1987).}

\lref\stein{H.~Stein, \ajou No\^{u}s &18 (84) 635.}

\lref\ak{A.~Kent, \ajou Int. J. Mod. Phys. &A5 (90) 1745.}

\lref\despagnat{B.~d'Espagnat, \ajou J. Stat. Phys. &56 (89) 747.}

\lref\samols{T.M.~Samols, {\it ``A Stochastic Model of a Quantum Field
Theory,''} Cambridge preprint DAMTP/94-39.}

\lref\sorkin{R.D.~Sorkin,
{\it ``Quantum Mechanics as Quantum Measure Theory,''} Syracuse preprint
SU-GP-93-12-1, gr-qc/9401003.}

\lref\bohm{D.~Bohm, \ajou Phys. Rev &85 (52) 166.}

\lref\grw{G.~Ghirardi, A.~Rimini and T.~Weber, \ajou Phys. Rev. &D34
(86) 470.}

\lref\gisin{N.~Gisin, \ajou Helv. Phys. Act. &62 (89) 363.}

\lref\percival{I.~Percival, {\it Primary State Diffusion}, QMW Physics
preprint 94 Apr 04.}

\lref\vonn{J.~von Neumann, {\it Mathematical Foundations of Quantum
Mechanics},
Princeton University Press, Princeton (1955).}

\lref\gosse{P.H.~Gosse, {\it ``Omphalos: an attempt to untie the geological
knot''}, (1857).}

\lref\wigner{E.P.~Wigner, {\it Remarks on the Mind-Body Question} in
{\it The Scientist Speculates}, I.J.~Good, ed., pp. 284-302,
Heinemann, London (1961).}


\Title{\vbox{\baselineskip12pt\hbox{ DAMTP/94-66}\hbox{ gr-qc/9409037}{}
}}
{\vbox{\centerline {Properties of Consistent Histories}}}{~}

\centerline{{\ticp Fay Dowker}}
\vskip.1in
\centerline{\sl Physics Department, University of
California-Santa Barbara,}
\centerline{\sl Santa Barbara, California 93106, U.S.A.}
\vskip.2in
\centerline{{\ticp Adrian Kent}}
\vskip.1in
\centerline{\sl Department of Applied Mathematics and
Theoretical Physics,}
\centerline{\sl University of Cambridge,}
\centerline{\sl Silver Street, Cambridge CB3 9EW, U.K.}

\bigskip

\centerline{\bf Abstract}
{We describe some properties of consistent sets of
histories in the Gell-Mann--Hartle formalism, and give an example 
to illustrate that one cannot recover the standard 
predictions, retrodictions and
inferences of quasiclassical physics using the criterion of
consistency alone. 
}

\newsec{Introduction}
\footline={\ifnum\pageno=0 {\hfil} \else\hss\tenrm\folio\hss \fi}

Standard quantum theory, in the Copenhagen 
interpretation,\foot{For 
definiteness consider the version of the 
Copenhagen interpretation set out by Landau and Lifschitz
\refs{\ll}.}
gives a robust and successful algorithm for predicting the 
results of laboratory experiments. It has, however, nothing to 
say about the larger quantum universe outside the lab, since its 
subject is solely the results of measurements made by 
some classical measuring apparatus whose existence is 
taken as an a priori assumption. 
New theories which make predictions without this last assumption would be
of great interest, since they would have greater predictive power.  

The consistent histories approach of Griffiths,
Omn\`es, Gell-Mann and Hartle has been suggested to be just such
a development, extending 
the Copenhagen interpretation \refs{\griff, \grifflogic, 
\omnes, \omnesreview} and allowing us to make predictions in 
quantum cosmology where the quantum system is the whole universe
\refs{\gmhsantafe, \gmhthree, \gmhprd}.   
It is a formalism from which, it is hoped, the largely
classical world of our experience might be {\it deduced}, rather 
than assumed. 
Thus, our observations of large-scale classical structure in the universe, 
of macroscopic objects following classical equations of motion, and
of definite classical outcomes to quantum experiments, are all supposed 
to be predictions, unconditionally derivable from the formalism.  
If these hopes were to be realised, the consistent histories 
approach would indeed have provided a significant increase  
in our predictive power.  
They rest, however, on as yet incomplete interpretational
arguments and have naturally led to much
debate \refs{\despagnat, \halliwell}.
Our own arguments, together with a critique of the existing literature,
will be given in detail elsewhere \refs{\dowkerkent}.
Our aim here is to draw attention to some perhaps counter-intuitive
properties of consistent sets of histories, most of which have not
previously been discussed in any detail in the consistent
histories literature, and to explain their physical relevance.

\newsec{The Consistent Histories Formalism}

We begin with a brief description of the consistent histories 
formalism as it applies to non-relativistic quantum mechanics, in the
Heisenberg picture, using the language of projection operators and density
matrices.
We assume that a Hilbert space, $\cal{H}$,
and hamiltonian, $H$,  are given, that hermitian operators correspond to
observables,
that the commutation relations amongst the hamiltonian and physically
interesting observables (such as position, momentum and spin) have
been specified, and
that the operators corresponding to the same observables at different times
are related by
\eqn\heis{
P (t) = \exp(i H t / \hbar) \, P(0) \, \exp( - i H t / \hbar ) \, .}
We are interested in a system
(in principle, the universe) whose initial density matrix $\rho_i$ is
given.  We require that $\rho_i$ is
positive semi-definite.  
The formalism also allows a final condition to be imposed, 
an interesting generalisation of standard quantum theory, though 
we shall not consider that possibility explicitly here.
The basic physical
events we are interested in correspond to sets $\sigma$ of orthogonal hermitian
projections $P^{(i)}$, with
\eqn\conds{\sum_i P^{(i)} = 1 \quad {\rm{and}}\quad
P^{(i)} P^{(j)} = \delta_{ij} P^{(i)}.}
These projective decompositions of the identity are  
applied at definite times, which we append to the sets
of projections: thus
$\sigma_j (t_j) =\{ P^{(i)}_j; i = 1,2,\ldots , n_j \}_{t_j}$. 
defines a set of projections obeying \conds\ and applied at time $t_j$.  
However, since our results depend only on the time ordering, 
we will generally omit explicit time labels. 
Suppose now we have a list of sets $\sigma_j (t_j)$ of this form, 
with $j$ running from $1$ to $n$, at times $t_j$
with $t_i<t_1<\ldots<t_n<t_f$.  
Then the histories given by choosing one projection from each
$\sigma_j$ in all possible ways are an exhaustive and exclusive set of
alternatives, $\cal S$.
We use Gell-Mann and Hartle's decoherence condition, and
say that the histories form a {\it consistent set} if
\eqn\decohgmh{ \Tr ( P^{(i_n )}_n \ldots P^{(i_1 )}_1
                               \rho_i P^{(j_1 )}_1
\ldots P^{(j_n )}_n ) =
\delta_{i_1 j_1 } \ldots \delta_{i_n j_n } p(i_1 \ldots i_n ) \, ,}
in which case $p(i_1 \ldots i_n )$ is the probability of the
history $P^{(i_1 )}_1 \ldots P^{(i_n )}_n$. (Gell-Mann and Hartle
term a set satisfying \decohgmh\ {\it{medium decoherent}}.)

We say the set
\eqn\extend{
\S' = (\rho_i ,\{\sigma_1,\dots, \sigma_k, \tau, \sigma_{k+1},\dots,
\sigma_n\})}
is a {\it consistent extension} of a consistent set of histories
$\S = (\rho_i, \{\sigma_1 , \ldots , \sigma_n \})$ by
the set of projections $\tau = \{Q^i : i = 1, \dots, m\}$ if $\tau$
satisfies \conds\ and $\S'$ is itself consistent.
We say the consistent extension $\S'$ is {\it trivial~} if, for
each history
$\{ P^{(i_1)}_1,\dots, P^{(i_k)}_k, P^{(i_{k+1})}_{k+1}, \dots, P^{(i_n)}_n\}$
in $\S$, at most one of the extended histories
$\{ P^{(i_1)}_1,\dots, P^{(i_k)}_k, Q^i, P^{(i_{k+1})}_{k+1},
\dots, P^{(i_n)}_n \}$ has non-zero probability.
We extend these definitions by taking 
consistent extension and trivial consistent extension
to be transitive relations.

\newsec{Counting consistent sets}

Let us now take the Hilbert space, ${\cal{H}}$,
to be of finite dimension $N$ and ask: just how 
many consistent sets are there? 
We first describe how consistent sets may be classified.
The basic objects in the formalism are the
projective decompositions of the identity
$\sigma_j =\{ P^{(i)}_j :  i = 1,2,\ldots , n_j\}$, where the
$P^{(i)}_j$ satisfy \conds.
These decompositions are parametrised by
the set of ranks $\{ r^{(1)}_j , r^{(2)}_j , \ldots , r^{(n_j )}_{j} \}$
of the projection operators, where
$N = \sum_{i=1}^{n_j} r^{(i)}_j$
and we take $r^{(1)}_j \geq r^{(2)}_j \geq \ldots \geq r^{(n_j )}_{j}$,
and the 
manifold:
\eqn\manifold{ G(N ; r^{(1)}_j , r^{(2)}_j , \ldots , r^{(n_j )}_{j} ) =
{{ U(N) } \over {( U( r^{(1)}_j )
\times U( r^{(2)}_j ) \times \ldots \times U( r^{(n_j )}_{j} ) )
\times J}} \, .}
$J$ is a discrete symmetry group that eliminates overcounting 
when some of the ranks are equal.

The parameter space of a set of histories is then a manifold,
$M$, 
which is a product of
such $G$'s, one for each projective decomposition, $\sigma_k$,
$k = 1,2,\dots,n$
(i.e. one
for each time, $t_k$): $M = G_1\times \dots \times G_n$.

It is easy to use this parametrisation in explicit calculations: one can
define projections $\{ P^{(1)} , \ldots , P^{(n_j )} \}$
of ranks $\{ r^{(1)} , r^{(2)} , \ldots , r^{(n_j )} \}$ by choosing an
orthonormal basis of vectors $\{ x_1 , \ldots , x_n \}$, so that
\eqn\projns{
P^{(1)} = \sum_{i=1}^{r_1} x_i (x_i )^{\dagger} \, , \quad
P^{(2)} = \sum_{i=r_1 +1}^{r_1 + r_2 } x_i (x_i )^{\dagger} \, ,
}
and so on.  The redundancies in this parametrisation correspond to the actions
of the quotient subgroups, and can be eliminated at any convenient
point. This can be done for each 
time. Thus, in principle, we can simply fix the form of the initial
density matrix,
fix the ranks of the projection operators in the type of consistent set we
wish to classify, and then impose the consistency conditions \decohgmh.
These will be algebraic equations defining  a submanifold, $L$,
the submanifold of consistent sets, of the
manifold of all sets of histories, $M$.

While these algebraic equations are generally very complicated, one
can at least make educated guesses at the
qualitative features of $L$, such as its 
dimension, and these guesses can
be checked in simple examples. 
A typical physical illustration of the consistent histories formalism
would involve a small number of projection operators, describing
quasiclassical operators in a large Hilbert space.
One might, for instance, describe a coarse-grained trajectory
of a dust grain, interacting with a photon background.
Such physical projection operators rarely form a precisely consistent
set, and there has been debate over whether or not it is possible
to find close approximations to the projection operators which are
exactly consistent.  
Comparing the number of parameters used to specify sets  
in $M$ (very large) with the number of consistency equations (rather 
small)
suggests
that this is generically possible. 
If so, there is no need to follow Gell-Mann and Hartle in ascribing
a fundamental role to approximately consistent sets: exactly
consistent sets suffice. Moreover, the counting arguments 
show that in any physically realistic
situation the dimension of $L$, the space of consistent sets, is
very large. 

\newsec{Properties of Consistent Sets}

So, let us suppose that physics is described by exactly consistent sets
and look at what this implies.  We omit proofs, which can be found
in Ref. \refs{\dowkerkent}.
\medskip\noindent
{\it{Lemma 1}}\foot{A similar result has been obtained
independently by Di\'osi \refs{\diosi}.} \qquad
Let $\S = ( \rho , \{ \sigma_1 , \ldots , \sigma_k \} )$ be a
consistent set which is not a trivial extension of any consistent
subset, defined on a space $\cal{H}$ of dimension $N$, with initial
density matrix $\rho$ of rank $r$.
Then the length $k$ of $\S$ obeys $k \leq r N $.
(In particular, if $\rho$ is pure then $k \leq N$.)
\medskip\noindent
In other words, if the Hilbert space of the 
universe is finite-dimensional there is a strict bound
on the number of probabilistic 
physical events. Once
this number has occurred, the evolution of the universe 
continues completely deterministically.
This is mathematically an unsurprising feature of the formalism but,
as far as we are aware, physically quite new: no previous
interpretation of quantum theory
has suggested that quantum stochasticity is exhaustible
in this way.

In the consistent histories approach, predictions can only 
be made once a consistent set ---
{\it the} physically relevant set --- is fixed.
The key problem in interpreting the formalism is explaining how,
given the profligate abundance of consistent sets,  
this is to be done. 
Once the choice has been made,
one can simply declare by fiat
that physics should be described by one history from the relevant set,
chosen at random using the decoherence functional probability
distribution. Again, the key question is whether the choice
has been made {\it within} the formalism or whether it 
relies on assumptions that 
go beyond it.

It thus becomes an important question 
whether, when some of the projective
decompositions in the relevant set are known, others can be
determined.
In particular, if, taking the past and present for
granted, we were able to
deduce the form of the relevant set in the
future using only consistency criteria then 
we could indeed make 
unconditional predictions about the future
within the consistent histories formalism: the choice
of set would be clearly determined by the formalism.  
This, though, is
generally false:
\medskip\noindent
{\it{Lemma 2}} \qquad Let $\S$ be a set of consistent histories
for which there exists a non-trivial consistent extension. 
Let $\S$ have a pure initial state $\rho$, and let ${\cal
H}$ be either finite-dimensional or separable.
Then there exists a continuous family of non-trivial extensions
for each history in $\S$ with non-zero probability.
\medskip\noindent
So, if a consistent set describing a physical system up to time $t$ leaves some
future events unpredictable, there are infinitely many different
consistent continuations of that set.  In particular, if a consistent
set describes, in Gell-Mann and Hartle's language, quasiclassical
physics --- involving operators describing the same types of variables
at different times, following largely deterministic evolution
equations --- up to time $t$ then, if any unpredictability remains,
almost all future consistent continuations will {\it not} be
quasiclassical.  Whatever our experience of a
persistently quasiclassical world may be ascribed to --- and there are
various suggestions \refs{\dowkerkent} --- it does not follow simply
from consistency. 

Still, one might hope that at least, if the set up to time $t$ is
quasiclassical, then any non-quasiclassical consistent future
extension can consistently incorporate future deterministic quasiclassical
predictions.  Indeed, Omn\`es has suggested that this
is so \refs{\omnesreview}.
But in fact, as Omn\`es now accepts, this fails quite
generally.  If unpredictability remains, then
there are no future predictions consistent with all consistent
extensions of the present data:
\medskip\noindent
{\it{Lemma 3}} \qquad Let $\S =
( \ket{\psi} \bra{\psi} , \{\sigma_1 , \ldots , \sigma_l \})$
be as in Lemma 2, with $\cal{H}$ finite-dimensional or separable.
Then there is no projective
decomposition $\sigma_{l+1}$ such that:\hfill\break
(i) $\S' = (\rho, \{\sigma_1 , \ldots , \sigma_l , \sigma_{l+1} \})$
is a consistent extension of $\S$\hfill\break
and (ii) any consistent extension
$\S'' = (\rho, \{\sigma_1 , \ldots , \sigma_l ,
\tau_{1} , \ldots , \tau_r \})$ of $\S$ has a consistent extension
$(\rho, \{\sigma_1 , \ldots , \sigma_l ,
\tau_{1} , \ldots , \tau_r , \sigma_{l+1} \})$.
\medskip\noindent
 We have only been able to identify one class of statements which
 {\it can} consistently be added to any consistent extension of a set.
These arise where the set contains the same decomposition twice.
In this case, further repetitions can be included, provided that they
are made between the first two:
\medskip\noindent
{\it{Lemma 4}} \qquad
Let $\S = ( \rho , \{ s_1 , \ldots , s_j , t , t_1 , \ldots , t_l , t,
s_{j+1}, \ldots , s_k \} ) \equiv (\rho, \{ S_1 , t, T , t, S_2 \} )$ be a
consistent set in which the projective decomposition $t$ is repeated.
Let $\S' = (\rho , \{ S_1 , t , T_1 , t , T_2 , t , S_2 \} )$ be an
extension of $\S$ by a further repetition of $t$ at some point between the
first two, so that $\{ T \} = \{ T_1 , T_2 \}$.
Then $\S'$ is also consistent.
\medskip\noindent
Put picturesquely, if a tree is observed standing in the forest at
dusk and dawn, and if the dynamics cause no qualitative complications,
then the formalism
allows us unambiguously to deduce that it remained standing
overnight while unobserved.

\newsec{Example} 

A simple example illustrates the weaknesses of the
consistency criterion.  
Consider two systems, described
by two-dimensional Hilbert spaces $V$ and $W$, with orthonormal 
bases $\{ v_1 , v_2 \}$ and $\{ w_1 , w_2 \}$.  We suppose that the 
total Hamiltonian is zero except between times 
$t=t_1$ and $t=t_2$, during which the systems are coupled by an
interaction which models a measurement process.  Specifically,
we take the unitary evolution operator between these times to be the
operator $U$ defined by
\eqn\evoln{
U \ket{ v_1 } \ket {w_1 }  = \ket{ v_1 } \ket {w_1 }  \, , 
U \ket{ v_2 } \ket {w_1 }  = \ket{ v_2 } \ket {w_1 }  \, , 
U \ket{ v_1 } \ket {w_2 }  = \ket{ v_2 } \ket {w_2 }  \, , 
U \ket{ v_2 } \ket {w_2 }  = -\ket{ v_1 } \ket {w_2 }  \, ,}
and we take $\rho_i = \ket{v_1 } \bra{v_1} \otimes I$.  
In this much simplified (and unrealistic) model, $V$ represents 
the relevant degrees of freedom --- two ``pointer positions'' ---
of a measuring device examining a
two-dimensional microscopic quantum system represented by $W$. 
We have chosen $\rho_i$ so that the initial pointer position 
is specified and no information is known about the system $W$,
which might for example be a spin-$1/2$ cosmic ray.

Now consider an experiment in which the pointer is observed to be in
state $\ket{v_1 }$ at time $t_3 > t_2$, so that the combined 
system lies in the range of the projection 
$P = \ket{v_1 } \bra{v_1} \otimes I$.
The standard description of this experiment would 
distinguish between the macroscopic system $V$, which follows
classical dynamics after $t_2$  and the microscopic system $W$, which
is observed at time $t_3$ to be in the state $\ket{w_1}$ 
and thereafter follows the Schr\"odinger equation.  
Now the consistent histories description of the observation uses
the set $\S$ defined by the single projective
decomposition $\{ P , 1 - P \}$ at time $t_3$, and
specifically the history from that set defined by $P$.

However, none of the standard inferences drawn from the observation can be
made using the consistency criterion alone: we cannot deduce that the
pointer was in state $\ket{v_1}$ at times between $t_2$ and $t_3$ or after
time $t_3$ (as classical mechanics would imply), nor that the 
system was in state $\ket{w_1}$ after $t_3$ (as standard quantum
mechanics would imply).  The reason is that in each case
we can find a consistent extension of $\S$ with which 
the inference is inconsistent.  

For example, 
let $Q$ be the projection $I \otimes  \ket{w } \bra{w}$ onto 
the state $\ket{w} = (\ket{w_1} + \ket{w_2} )/ \sqrt{2}$ and 
$R$ the projection $I \otimes  \ket{w_1 } \bra{w_1}$.
Then the set $\{  \{ P , 1 - P \} (t_3 ) , \{ Q , 1-Q \} (t ) \}$ 
is consistent.  
(We now include explicit time labels and take $t_3 < t < t'$.) 
However, since
the extended
set $\{  \{ P , 1 - P \} (t_3 ) , \{ Q , 1-Q \} (t) , \{ R , 1-R \}(t' ) \}$
is inconsistent, we cannot infer that
the quantum mechanical system $W$ is in state $\ket{w_1}$ at any time
$t' > t_3$.  Essentially the same argument shows that this inference 
cannot be made at time $t_3$.  In other words, the observation of a 
pointer state does not imply the result of its measurement in the
usual way.  Similar arguments show that we
cannot infer that the pointer itself is in state $\ket{v_1}$ at any time
other than $t_3$.

\newsec{Conclusions}

These results illustrate interesting features of the consistent
histories formalism.  Those who
prefer their fundamental theories to be mathematically precise will
be encouraged that the use of approximately consistent sets can
apparently be avoided.
On the other hand, it will be seen that 
consistency itself is a very weak condition: there is generally a 
large variety of consistent sets. 
We stress that this does not mean that the consistency criterion
is incompatible with experiment.
Many careful studies have illustrated 
the efficiency of the 
decoherence process and its crucial importance in understanding 
the dynamics of quasiclassical systems.\foot{
See for example the decoherence calculations of ref. \refs{\jooszeh}.}
Assuming that quantum theory holds good as a description of
macroscopic systems, the moral to be drawn from these studies is that,
in any realistic description of an experiment, one of the consistent sets will 
correctly describe familiar quasiclassical 
physics and will allow the standard predictions and inferences.  

The problem here is that all consistent sets have the same status in 
the formalism, most have very little to do with the 
quasiclassical world of our observations, and we 
can only make the predictions we would like to make after we
have made the choice of one particular set: the familiar
quasiclassical one.   
If the consistent histories formalism is to
represent an enhancement in predictive 
power over the Copenhagen interpretation it
requires interpretational arguments that show this choice
to be determined by the formalism. 
Omn\`es' interesting attempt\refs{\omnes, \omnesreview} to find  
a mathematical criterion that correctly 
identifies the relevant set, unfortunately, fails. 
The remaining arguments in the
literature \refs{\gmhsantafe, \gmhthree}
which suggest that the formalism nonetheless does explain the
observed persistence of quasiclassicality therefore deserve careful 
scrutiny.  Our own
conclusion \refs{\dowkerkent} is that they rely on important 
hidden assumptions.  

\bigskip\bigskip\centerline{{\bf Acknowledgements}}\nobreak
We thank the Aspen Institute for their hospitality
in the formative stages of this work. F.D. thanks, in turn,
the Newton Institute in Cambridge for {\it their}
hospitality during part of
this work.
F.D. was supported by NSF grant no. PHY-9008502;
A.K. by a Royal Society University Research
Fellowship.
It is a pleasure to thank Murray Gell-Mann, Robert Griffiths and
Roland Omn\`es for invaluable discussions of their work.
We are particularly grateful to Jim Hartle for many helpful and
stimulating discussions.
\listrefs
\end